\documentclass[10pt]{article}

\usepackage[margin=1in]{geometry}
\usepackage{times}
\usepackage{microtype}
\usepackage{graphicx}
\usepackage{booktabs}
\usepackage{amsmath,amssymb}
\usepackage{url}
\usepackage{hyperref}
\usepackage{enumitem}
\usepackage{algorithm}
\usepackage{algpseudocode}
\usepackage{xcolor}
\usepackage{comment}

\title{\vspace{-0.8em}StepCache: Step-Level Reuse with Lightweight Verification and Selective Patching for LLM Serving\vspace{-0.5em}}
\author{
Azam Nouri\\
\small Department of Science, Technology \& Mathematics, Lincoln University, USA
}
\date{}

\begin{document}
\maketitle

\begin{abstract}
We address LLM serving workloads where repeated requests share a common solution structure but differ in localized constraints (e.g., output schema, variable names, or numeric constants). Prior caching approaches typically reuse either full responses (semantic caching) or model-internal KV/prefix states, which are respectively brittle under partial changes or tightly coupled to specific backends. We present \textbf{StepCache}, a backend-agnostic step-level reuse layer that segments outputs into ordered steps, retrieves the best-matching cached request, verifies steps using lightweight \emph{task-aware} checks, and regenerates only failing regions via selective patching. StepCache additionally supports strict structured-output enforcement for JSON (single-step extraction, required-key constraints, one-shot repair) and conservative \emph{skip-reuse} fallbacks for semantic changes. For linear equations, StepCache promotes verification into correction via a bounded repair loop with a deterministic fallback that guarantees correctness when the backend model fails.

In a CPU-only perturbation-heavy micro-benchmark on math and JSON variants (10 base prompts per task, $k{=}3$ variants per perturbation; 222 evaluation requests per seed), averaged over three seeds (42--44), StepCache reduces mean latency from 2.13s to 0.67s (median: 2.42s to 0.01s; p95: 3.38s to 3.30s), reduces total token usage from 36.1k to 27.3k (162.7$\rightarrow$123 tokens/request), and improves end-to-end correctness from 72.5\% to 100\% under both task-specific checks and a stitched-output integrity check. Across requests, 79.7\% take the reuse-only fast path, 5.4\% require patching, and 14.9\% trigger skip-reuse. 
\end{abstract}

\section{Introduction}
Large language models (LLMs) are increasingly deployed in interactive systems where end-to-end latency and service cost are first-order constraints (e.g., code assistants, agentic workflows, and structured-data generation). Recent LLM serving work has largely focused on backend efficiency---notably GPU memory management and scheduling for the KV cache---and on caching strategies that avoid recomputing outputs for repeated or highly similar requests.

Despite this progress, existing caching mechanisms typically operate at coarse granularity and behave as \emph{all-or-nothing} reuse:
\begin{itemize}[leftmargin=1.2em]
\item \textbf{Semantic response caching} returns a previously generated \emph{entire response} when a new prompt is sufficiently similar in the embedding space, sometimes augmented with verification or confidence thresholds.
\item \textbf{Prefix/KV caching} reuses \emph{model-internal states} for repeated prefixes but is tied to a particular model, tokenizer, and context, and primarily accelerates repeated prompt prefixes rather than reuse of parts of an answer.
\end{itemize}

In many real workloads, requests are only \emph{partially} similar: the solution structure is stable, but a small localized change is required due to a new constraint or parameter. Examples include adding a required key to a JSON schema (``add key \texttt{d}''), changing the target variable (solve for $y$ instead of $x$), or modifying constants in a familiar template (e.g., changing $2x+3=c$). In such cases, whole-response reuse can be incorrect, while full regeneration recomputes large portions of an otherwise reusable solution. This gap motivates a more granular reuse unit that can preserve correctness under localized changes.

We present \textbf{StepCache}, a backend-agnostic reuse layer that treats an answer as an ordered sequence of \emph{steps} and supports \emph{selective regeneration}. StepCache segments prior outputs into steps, retrieves a best-matching cached request, verifies cached steps with lightweight \emph{task-aware} checks, and regenerates only the failing region via contiguous block patching. Because StepCache operates above the model runtime and does not depend on model-internal KV states, it can be placed in front of existing serving engines as a drop-in optimization across backends.

At the same time, StepCache's strongest correctness guarantees arise in settings where inexpensive verifiers exist (e.g., parsing structured JSON with required keys, or checking linear-equation consistency). For open-ended generations without checkable invariants, verification is necessarily weaker or more costly (e.g., using an LLM judge), and StepCache may revert to conservative policies such as \emph{skip-reuse}. Designing broadly applicable, low-cost verifiers for open-ended text is an important direction for future work. 

\paragraph{Contributions.}
\begin{itemize}[leftmargin=1.2em]
\item \textbf{Step-level reuse.} We introduce a step-granular caching abstraction for LLM serving: retrieve a single best-matching prior request, preserve step order, verify per-step, and selectively regenerate only failing regions rather than whole responses.
\item \textbf{Selective patching with safe fallbacks.} We develop contiguous block patching for dependency-heavy reasoning and an \emph{adaptive skip-reuse} policy that falls back to full regeneration when inconsistency signals predict that reuse would be unproductive.
\item \textbf{Task-aware verification and repair.} We implement lightweight verifiers and bounded repair loops for two representative families: structured JSON generation (single-step extraction, required-key constraints, one-shot repair) and linear-equation solving (math consistency checks, variable-aware validation, one-shot repair).
\item \textbf{Instrumented prototype and evaluation.} We provide an end-to-end prototype with detailed accounting (calls, tokens, latency) and a perturbation-heavy micro-benchmark that quantifies reuse/patch rates, robustness, and tail behavior.
\end{itemize}

\section{Background and Related Work}
\paragraph{LLM serving and KV-cache management.}
A large body of work improves LLM serving efficiency by optimizing the model runtime, especially memory management and scheduling of the attention KV cache. PagedAttention (as used in vLLM) reduces fragmentation and enables efficient paging of KV states for high-throughput decoding \cite{kwon2023pagedattention}. More recent systems further reduce KV overhead via compression and eviction policies \cite{feng2025evicpress} or reconfigure KV allocation to meet tail-latency/SLO targets under long-context workloads \cite{ma2026orbitflow}. StepCache is complementary to these approaches: it operates \emph{above} the serving backend as an application-layer reuse mechanism (step retrieval, verification, and selective regeneration), and can be composed with KV-cache optimizations to reduce both end-to-end latency and backend compute when repeated requests share partial structure.

\paragraph{Workload characteristics and tail latency.}
Real serving workloads exhibit burstiness, skew, and heavy-tailed latency distributions. BurstGPT provides traces and analysis that emphasize the importance of tail metrics and robustness under load \cite{wang2024burstgpt}. Our evaluation reports mean, median, and p95 latency and explicitly distinguishes the fast-path (reuse) from the slow-path (patching or skip-reuse fallback). Integrating StepCache with trace replay on public workloads is a natural next step.

\paragraph{Semantic response caching and verification.}
Semantic caching systems accelerate applications by returning a previously generated response when a new prompt is sufficiently similar, often using embedding retrieval with optional verification. GPTCache is a representative open-source semantic cache \cite{bang2023gptcache}, and recent work proposes verified semantic caching mechanisms that aim to provide correctness guarantees when reusing cached outputs (e.g., vCache \cite{schroeder2025vcache} and tiered verified caching \cite{singh2026asynchronous}). These approaches typically treat the \emph{entire response} as the reuse unit and therefore make an all-or-nothing decision: either the cached response is accepted as-is, or the system regenerates a new response. StepCache targets a complementary regime where prompts share a stable solution skeleton but differ in localized constraints. It uses \emph{ordered steps} as the reuse unit, performs \emph{per-step} lightweight verification, and applies \emph{contiguous block patching} to regenerate only the first failing step and its downstream dependents; when inconsistency signals suggest reuse is unlikely to be beneficial, StepCache conservatively triggers \emph{skip-reuse} and falls back to full regeneration.

\section{StepCache Design}
\subsection{Overview}
\begin{figure}[t]
  \centering
  \includegraphics[
        height=0.45\textheight,
        keepaspectratio
    ]{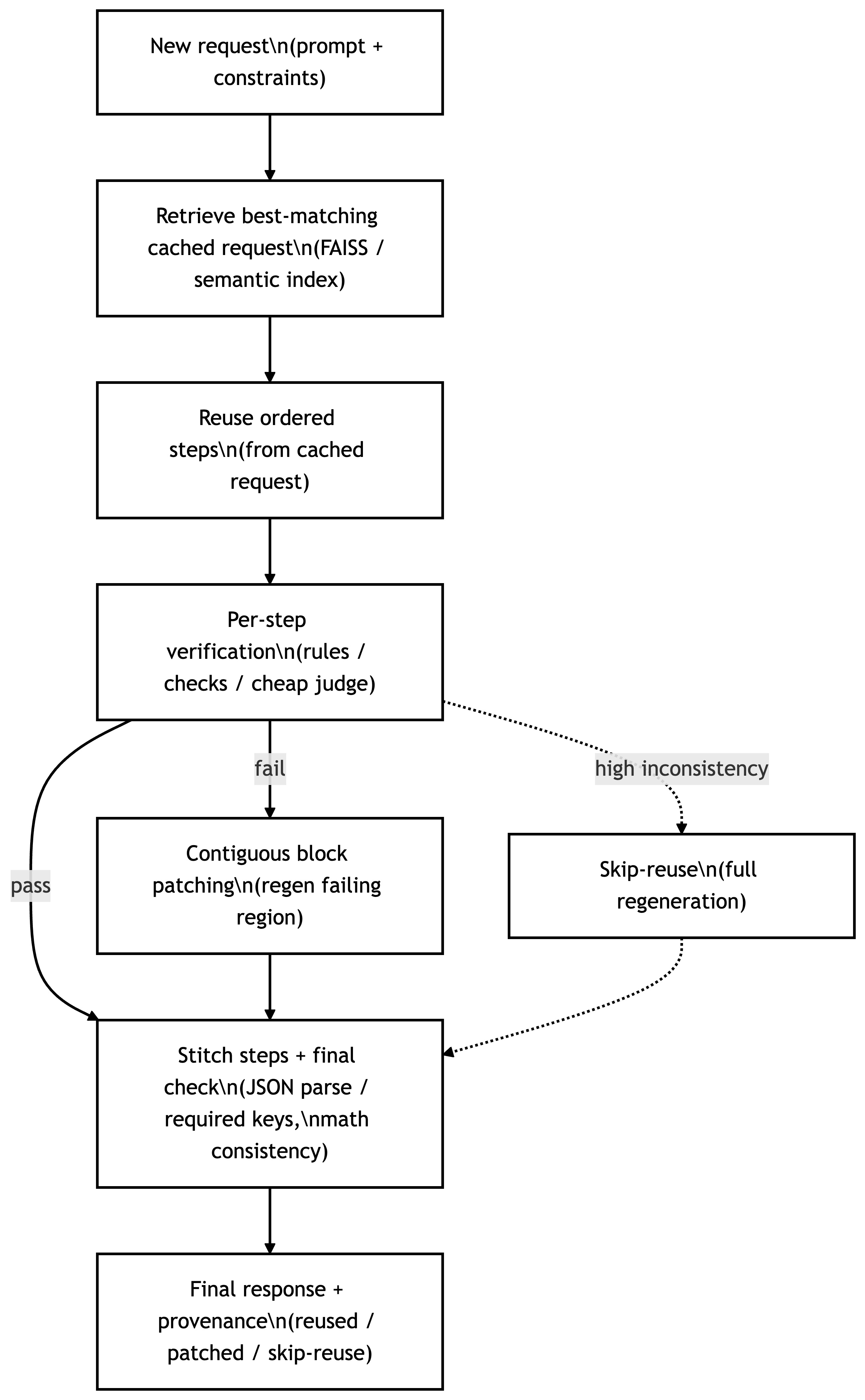}
  \caption{StepCache pipeline: retrieval, per-step verification, selective regeneration via contiguous block patching, and stitched-output integrity checks.}
  \label{fig:pipeline}
\end{figure}

Figure~\ref{fig:overview} provides a compact conceptual view of StepCache, while Figure~\ref{fig:pipeline} shows the end-to-end pipeline. On the first occurrence of a request (or during a warmup phase), StepCache forwards the prompt to the backend LLM, segments the generated output into an ordered list of steps, and stores (i) a prompt embedding for retrieval and (ii) the step sequence with lightweight metadata (task type and constraints).

For a new request, StepCache computes the prompt embedding, retrieves the single best-matching cached request (prompt-to-prompt similarity), and evaluates the cached step sequence under the new prompt and constraints. For each step, StepCache:
\begin{itemize}[leftmargin=1.2em]
\item \textbf{reuses} the step if it passes verification;
\item \textbf{patches} the step if it fails verification by regenerating only the failing region, typically as a contiguous block to respect step dependencies;
\item \textbf{falls back} to full regeneration (\emph{skip-reuse}) when inconsistency signals indicate that reuse is likely to require extensive patching or reduce correctness.
\end{itemize}

StepCache then stitches the resulting step list into the final response and applies a task-level integrity check (e.g., JSON parse + required keys; math solution consistency). When the stitched output fails this final check, StepCache executes a bounded task-specific repair (one-shot for JSON and math in our prototype) and revalidates before returning the answer.

\begin{figure}[t]
\centering
\fbox{\begin{minipage}{0.95\linewidth}
\small
\textbf{StepCache pipeline:}\\
(1) \textbf{Embed} prompt $\rightarrow$ (2) \textbf{Retrieve} best cached request $\rightarrow$ (3) \textbf{Verify} each cached step \\
$\rightarrow$ (4) \textbf{Reuse} PASS steps + \textbf{Patch} FAIL steps (block) or \textbf{Skip-reuse} $\rightarrow$ (5) \textbf{Stitch} \\
$\rightarrow$ (6) \textbf{Final checks} (math/JSON) + bounded \textbf{repair} $\rightarrow$ Answer + per-step provenance.
\end{minipage}}
\vspace{-0.3em}
\caption{StepCache overview. StepCache reuses verified steps and selectively regenerates only failing steps.}
\label{fig:overview}
\vspace{-0.8em}
\end{figure}

\subsection{Caching unit: steps}
StepCache treats a response as an ordered sequence of \emph{steps}. By default, segmentation is heuristic and task-agnostic: we split on paragraph boundaries (double newlines), explicit enumerations (e.g., ``Step 1''), and list delimiters. For structured-output tasks, segmentation becomes task-aware. In particular, for JSON generation we enforce \textbf{single-step segmentation}: StepCache extracts the first syntactically valid JSON object/array from the model output (removing code fences and surrounding text when present) and caches that JSON payload as the sole step. This design ensures that downstream verification and patching operate on a single structured unit.

For each cached request, StepCache stores:
\begin{itemize}[leftmargin=1.2em]
\item a \textbf{prompt embedding} used for prompt-to-prompt retrieval,
\item the \textbf{ordered step texts} produced for that request,
\item \textbf{constraints metadata} (e.g., task type, required keys),
\item optional \textbf{tool outputs} (not exercised in our micro-benchmark),
\item \textbf{provenance and timing} signals used for accounting (reuse/patch/skip decisions).
\end{itemize}

\subsection{Retrieval}
In its MVP configuration, StepCache retrieves the \emph{single} best-matching cached request by prompt embedding similarity and reuses its step list in the original order. To isolate the effect of step-level reuse, we do not mix steps across different cached requests. We implement approximate nearest-neighbor retrieval with FAISS and store per-request metadata in a local database \cite{johnson2017faiss}.

\subsection{Verification}
\label{subsec:verification}
StepCache uses lightweight task-aware verification to decide whether cached steps remain valid under the new prompt and constraints. Our prototype implements a \textbf{rule-based verifier} for two representative task families.

\paragraph{Math (linear equations).}
For prompts of the form $a\cdot v + b = c$ (where $v$ is the target variable), StepCache parses $(a,b,c,v)$, computes the expected solution $v^\star = (c-b)/a$, and flags cached steps that contradict these values. The verifier checks for:
\begin{itemize}[leftmargin=1.2em]
\item incorrect final assignments (e.g., $v=5$ when $v^\star=6$),
\item incorrect intermediate equalities (e.g., $a v = N$ where $N \neq c-b$),
\item incorrect stated equation constants (e.g., $a v + b = N$ where $N \neq c$).
\end{itemize}
If the first inconsistency occurs at step index $i$, StepCache conservatively marks the suffix of steps $i..$end as failing, enabling contiguous block patching that respects step dependencies.

\paragraph{JSON (required keys).}
StepCache validates cached JSON by parsing it and optionally enforcing a \texttt{required\_keys} constraint. A step fails verification if parsing fails or if any required key is missing. This verifier integrates with a strict patch prompt and a bounded repair attempt for structured-output recovery.

\subsection{Selective patching and repair}
Algorithm~\ref{alg:stepcache} sketches StepCache inference. Given a retrieved step sequence, StepCache first identifies failing steps via task-aware verification (Section~\ref{subsec:verification}, where applicable). Rather than regenerating the entire response, StepCache performs \emph{selective regeneration}: it reuses verified steps and regenerates only the minimal region required to restore correctness under the new prompt and constraints.

When failures are detected, StepCache applies one of two patching strategies depending on task structure:
\begin{itemize}[leftmargin=1.2em]
\item \textbf{Contiguous block patching (math).} For dependency-heavy reasoning such as equation solving, StepCache patches a suffix of steps starting at the first inconsistent step index $i$ through the end. Each patch call includes a \texttt{math\_state\_hint} containing $(a,b,c,v,v^\star,c-b)$ (variable name, expected solution, and expected intermediate value) to prevent reuse of stale constants and to constrain the regenerated steps to be numerically consistent.
\item \textbf{Strict structured patching (JSON).} For JSON tasks, StepCache patches the (single) structured step using a dedicated prompt that requires the output to be \emph{valid JSON only} (no markdown or explanations), enforces any \texttt{required\_keys} constraints, and provides a schema example. After patching, StepCache re-parses the output and checks required keys; if validation fails, it performs at most one additional \emph{repair} attempt using error feedback (parse error or missing keys).
\end{itemize}

\paragraph{Conservative skip-reuse and deterministic fallback (math).}
Selective patching is not always cost-effective under semantic changes that invalidate most of the cached solution (e.g., changing constants in an equation). StepCache therefore uses a conservative \emph{skip-reuse} policy for math: (i) if the parsed equation state $(a,b,c,v)$ differs between the new prompt and the retrieved cached request, StepCache bypasses reuse; (ii) if the first inconsistent step is step 1 (i.e., no cached step verified) or the fraction of inconsistent steps exceeds a threshold (0.5 in our prototype), StepCache also bypasses reuse. In our benchmark we additionally mark \texttt{value\_change} cases with a \texttt{force\_skip\_reuse} constraint to isolate semantic-change behavior.

To guarantee correctness for this checkable task family, StepCache applies a bounded repair attempt and, if checks still fail, returns a minimal deterministic solution of the form ``$v = v^\star$'' computed from the parsed equation. This turns inexpensive verification into a correctness-preserving fallback for linear equations.

\begin{algorithm}[t]
\caption{StepCache inference (simplified; 1-indexed steps)}
\label{alg:stepcache}
\begin{algorithmic}[1]
\Require prompt $p$, task type $t$, constraints $C$, cache $\mathcal{D}$
\State $e \leftarrow \mathrm{Embed}(p)$
\State $(p', S) \leftarrow \mathrm{RetrieveBest}(\mathcal{D}, e)$ \Comment{$S=[s_1,\dots,s_m]$}
\If{$S$ is empty} \Comment{cache miss}
  \State $A \leftarrow \mathrm{LLMGenerate}(p)$; $S \leftarrow \mathrm{Segment}(A)$; store in $\mathcal{D}$; \Return $A$
\EndIf
\If{$t=\textsc{Math}$}
  \If{$C$ indicates \textsc{ForceSkip} \textbf{or} $\mathrm{MathState}(p) \neq \mathrm{MathState}(p')$}
     \State \Return $\mathrm{LLMGenerate}(p)$ \Comment{conservative skip-reuse on semantic change}
  \EndIf
  \State $i \leftarrow \mathrm{FirstInconsistentIndex}(S, p)$ \Comment{$i \in \{1,\dots,m\}$ or $\bot$}
  \If{$i \neq \bot$ and ($i=1$ or $\mathrm{InconsistentFrac}(S,p)\ge 0.5$)}
     \State \Return $\mathrm{LLMGenerate}(p)$ \Comment{skip-reuse}
  \ElsIf{$i \neq \bot$}
     \State mark steps $i..m$ as FAIL
  \EndIf
\EndIf
\For{$j=1$ to $m$}
  \If{$s_j$ is not marked FAIL}
    \State $v_j \leftarrow \mathrm{Verify}(s_j, p, t, C)$
    \If{$v_j=\textsc{FAIL}$} mark $j$ as FAIL \EndIf
  \EndIf
\EndFor
\State $S' \leftarrow$ reuse PASS steps; patch FAIL steps (block/structured)
\State $A' \leftarrow \mathrm{Stitch}(S')$
\If{$\mathrm{FinalCheck}(A',p,t,C)=\textsc{FAIL}$}
  \State perform at most 1 task-specific repair and recheck
  \If{still FAIL and $t=\textsc{Math}$}
     \State \Return $\mathrm{DeterministicSolve}(p)$
  \EndIf
\EndIf
\State \Return $A'$
\end{algorithmic}
\end{algorithm}

\section{Implementation}
We implement StepCache as a thin Python layer in front of an OpenAI-compatible chat-completions API. The design is intentionally \emph{backend-agnostic}: StepCache does not access model-internal KV states and requires only standard request/response I/O plus token usage metadata when available. This allows StepCache to sit in front of GPU serving engines (e.g., vLLM) or local/remote model endpoints without modifying the model runtime.

For math, StepCache uses robust prompt parsing to detect semantic changes in $(a,b,c,v)$ and conservatively triggers skip-reuse; if task checks still fail after a bounded repair attempt, it returns a minimal deterministic solution ``$v=v^\star$'' computed from the parsed equation.

\paragraph{Components.}
\begin{itemize}[leftmargin=1.2em]
\item \textbf{Embeddings.} We compute prompt embeddings using SentenceTransformers \texttt{all-MiniLM-L6-v2}, based on Sentence-BERT style bi-encoders and MiniLM distillation \cite{reimers2019sbert,wang2020minilm}.
\item \textbf{Index and storage.} We perform approximate nearest-neighbor retrieval with a FAISS index and store per-request metadata (step lists, task constraints, counters) in a local database \cite{johnson2017faiss}.
\item \textbf{Backend model.} For our micro-benchmark, we use a small local model served through an OpenAI-compatible endpoint (Qwen2.5-3B in our experiments) \cite{yang2024qwen25}. Swapping the backend to a GPU engine requires no changes to StepCache \emph{semantics}. 

\item \textbf{Instrumentation and accounting.} StepCache performs structured logging of every backend call and maintains explicit counters for baseline generation, cache-hit reuse, verification, patch calls, and skip-reuse fallbacks. The benchmark records per-request latency and token usage and reports both mean and tail statistics (median, p95), enabling clear separation of fast-path reuse from slow-path patching/fallback behavior.
\end{itemize}

\section{Evaluation}

\subsection{Experimental setup}
We evaluate StepCache using a CPU-only micro-benchmark designed to stress \emph{perturbation robustness}: each base task is instantiated into multiple variants that preserve overall intent while modifying local constraints. We include two representative task families:
\begin{itemize}[leftmargin=1.2em]
\item \textbf{Math (linear equations).} Solve prompts of the form $a\cdot v + b = c$ (e.g., $2x+3=13$) under paraphrases and under a semantic perturbation that changes the right-hand-side constant (\texttt{value\_change}).
\item \textbf{JSON (structured output).} Produce a JSON object with specified keys under paraphrases and a constraint perturbation that adds a required key (\texttt{keys\_change}).
\end{itemize}
Perturbation levels include low/medium/high paraphrases (and constraint additions), plus \texttt{value\_change} and \texttt{keys\_change}. Each run consists of a warmup phase that seeds the cache and an evaluation phase that measures retrieval, verification, patching, and fallback behavior.

We compare:
\begin{itemize}[leftmargin=1.2em]
\item \textbf{Baseline:} call the backend model for each request.
\item \textbf{StepCache:} warmup seeds the cache; evaluation uses retrieval + verification + selective patching, with adaptive skip-reuse fallbacks when triggered.
\end{itemize}

\paragraph{Benchmark protocol.}
We construct a perturbation-heavy micro-benchmark from $n$ base prompts (e.g., one math template and one JSON template), and generate $k$ perturbed variants per base prompt across perturbation types (e.g., low/med/high and constraint changes). We use a two-phase protocol: a warmup phase that forces generation to seed the cache for each base template, followed by an evaluation phase that measures reuse, patching, skip-reuse, tokens, and latency on the perturbed requests. In our main run, we use $n{=}10$ base prompts per task and $k{=}3$ variants per perturbation, yielding 222 evaluation requests per seed (seeds 42--44). We report mean/median/p95 latency, token usage, outcome split (reuse-only/patch/skip), and correctness under both task-specific checks and stitched-output integrity checks. -

\subsection{Metrics}
We report reuse, patch, and skip-reuse rates; backend call counts; token usage; and latency statistics (mean/median/p95). Correctness is measured by (i) a task-specific quality check (math solution consistency; JSON parse and required keys) and (ii) a stitched-output integrity check applied to the final response, with recorded failure reasons (to detect false positives from step-level checks).


\subsection{Results}
Table~\ref{tab:summary} reports seed-averaged results over 222 evaluation requests per seed (seeds 42--44). StepCache reduces mean latency from 2.13s to 0.67s while maintaining perfect correctness under both checks (100\% quality and 100\% final-check pass). Median latency drops to near-zero (0.01s) because most requests take the reuse-only fast path; tail latency remains dominated by the slow path (patching or skip-reuse), yielding a modest change in p95 latency. Overall token usage decreases by $\sim$24\% (36.1k$\rightarrow$27.3k), reflecting that most requests avoid full regeneration and that patch prompts are kept compact.

\begin{table}[t]
\centering
\small
\begin{tabular}{lrr}
\toprule
Metric & Baseline & StepCache \\
\midrule
Mean latency (s)   & 2.13$\pm$0.04 & 0.67$\pm$0.04 \\
Median latency (s) & 2.42$\pm$0.05 & 0.01$\pm$0.00 \\
p95 latency (s)    & 3.38$\pm$0.03 & 3.30$\pm$0.04 \\
Total tokens (all) & 36.1$\pm$0.1k & 27.3$\pm$0.9k \\
Tokens / request   & 162.7$\pm$0.6 & 123.0$\pm$4.0 \\
Quality pass rate  & 72.5$\pm$0.5\% & 100\% \\
Final-check pass rate & 72.5$\pm$0.5\% & 100\% \\
Outcome split (Reuse-only / Patch / Skip) & -- & 79.7$\pm$2.4\% / 5.4\% / 14.9$\pm$2.4\% \\
\bottomrule
\end{tabular}
\vspace{-0.4em}
\caption{CPU-only micro-benchmark summary averaged over three seeds (42--44). Each seed uses 222 evaluation requests ($n{=}10$ base prompts per task, $k{=}3$ variants per perturbation). StepCache reduces mean latency by $\sim$3.2$\times$ and reduces total token usage by $\sim$24\%, while improving correctness from $\sim$72.5\% to 100\% under both task-specific checks and stitched-output integrity checks.}
\label{tab:summary}
\vspace{-0.9em}
\end{table}

Table~\ref{tab:perturb} provides a breakdown by task and perturbation. The \texttt{keys\_change} setting forces regeneration of the structured JSON step (100\% patch) but remains correct with modest overhead. The semantic-change setting for math (\texttt{value\_change}) triggers skip-reuse by construction, avoiding unproductive reuse under changed equation constants. For linear equations, StepCache additionally guarantees correctness via a bounded repair attempt and a deterministic fallback when checks fail.

\begin{table}[t]
\centering
\small
\begin{tabular}{llrrrrr}
\toprule
Task & Perturb & ReuseOnly\% & Patch\% & Skip\% & Tokens Saved & Final\% \\
\midrule
json & low          & 100.0 & 0.0   & 0.0   & 119  & 100 \\
json & med          & 100.0 & 0.0   & 0.0   & 84   & 100 \\
json & high         & 100.0 & 0.0   & 0.0   & 89   & 100 \\
json & keys\_change & 0.0   & 100.0 & 0.0   & -42  & 100 \\
\midrule
math & low          & 96.7  & 0.0   & 3.3   & -17  & 100 \\
math & med          & 96.7  & 0.0   & 3.3   & 19   & 100 \\
math & high         & 96.7  & 0.0   & 3.3   & 18   & 100 \\
math & value\_change & 0.0  & 0.0   & 100.0 & -2   & 100 \\
\bottomrule
\end{tabular}
\vspace{-0.4em}
\caption{Per-task breakdown averaged over seeds 42--44. ReuseOnly/Patch/Skip are mutually exclusive per-request outcomes. \texttt{keys\_change} forces structured patching for JSON. \texttt{value\_change} is treated as a semantic-change case and triggers skip-reuse. Tokens Saved is the per-request average difference in token usage (baseline minus StepCache), rounded to the nearest integer.}
\label{tab:perturb}
\vspace{-0.8em}
\end{table}

\section{Discussion}
\paragraph{Where StepCache provides benefit.}
StepCache provides the largest gains when the solution structure is stable across requests and perturbations are localized (e.g., paraphrases, formatting constraints, or small schema edits). In these settings, verification passes and the system executes the fast path (reuse), yielding large latency and token reductions.

\paragraph{Semantic changes and adaptive policies.}
When a perturbation changes core semantics (e.g., changing the constants in an equation), naive reuse can require patching most steps and may be less efficient than full regeneration. StepCache makes this behavior explicit through an adaptive skip-reuse policy driven by inconsistency signals, trading a small number of fallback backend calls for stable correctness and improved tail behavior.

\paragraph{Integrity considerations.}
Any caching layer introduces potential integrity risks (e.g., over-aggressive reuse under adversarial similarity, cache poisoning, or stale step reuse under unmodeled constraints). StepCache reduces these risks with task-aware verification, conservative regeneration policies, and stitched-output validation. Incorporating stronger verifiers (e.g., LLM-as-judge or formal checkers), access controls, and cache hardening are important directions for future work.

\section{Conclusion}
We introduced \textbf{StepCache}, a backend-agnostic caching layer for LLM serving that operates on \emph{step-level} reuse rather than whole responses or model-internal KV states. StepCache retrieves a best-matching prior request, verifies cached steps with lightweight task-aware checks, and selectively regenerates only failing regions via contiguous block patching, with bounded repair loops for structured outputs and a conservative skip-reuse fallback when reuse is predicted to be unproductive.

In a CPU-only micro-benchmark of math and JSON tasks with perturbation-heavy variants (222 evaluation requests per seed; $n{=}10$, $k{=}3$), averaged over three seeds (42--44), StepCache reduces mean latency from 2.13s to 0.67s (median: 2.42s to 0.01s; p95: 3.38s to 3.30s) and reduces total token usage from 36.1k to 27.3k (162.7$\rightarrow$123 tokens/request), while improving correctness from 72.5\% to 100\% under both task-specific checks and a stitched-output integrity check. Across requests, 79.7\% take the reuse-only fast path, 5.4\% require patching, and 14.9\% trigger skip-reuse under semantic change or inconsistency signals.

Future work includes (i) GPU-backed evaluation in front of production serving engines (e.g., vLLM) with throughput and utilization metrics, (ii) evaluation under realistic bursty traces (e.g., BurstGPT-style replay), (iii) extending verifiers beyond math/JSON to code and tool-augmented agent workflows, and (iv) cache integrity hardening against adversarial similarity and poisoning.

\vspace{-0.3em}
\section*{Reproducibility}
Our prototype is instrumented to log per-request decisions and costs, including reuse/patch/skip-reuse outcomes, backend call types, token usage, and latency breakdowns. Each run produces (i) a machine-readable \texttt{benchmark\_results.json} containing per-request records and aggregate statistics (mean/median/p95), and (ii) a \texttt{benchmark\_mismatches.json} file that captures cases where task-level checks and stitched-output checks disagree (including a failure reason), enabling targeted debugging of false positives/negatives.

To reproduce the reported results, run the benchmark script with a fixed seed and the desired cache configuration (e.g., \texttt{verify\_patch} mode) and disable optional code tasks:

\begin{verbatim}
python scripts/benchmark_perturb.py -n 10 -k 3 --seed 42 --include-code 0
python scripts/benchmark_perturb.py -n 10 -k 3 --seed 43 --include-code 0
python scripts/benchmark_perturb.py -n 10 -k 3 --seed 44 --include-code 0
\end{verbatim}

\cite{nouri2025ijca}
\cite{nouri2025sobel}
\cite{nouri2026siggain}

\bibliographystyle{plain}

\end{document}